# EVOLUTION OF NATURAL PATTERNS FROM RANDOM FIELDS

Lovrenc ŠVEGL[1] & Igor GRABEC[2]
[1] Faculty of Natural Sciences and Engineering,
[2] LASIN - Faculty of Mechanical Engineering, University of Ljubljana, Slovenia

**Abstract:** In the article a transition from pattern evolution equation of reaction-diffusion type to a cellular automaton (CA) is described. The applicability of CA is demonstrated by generating patterns of complex irregular structure on a hexagonal and quadratic lattice. With this aim a random initial field is transformed by a sequence of CA actions into a new pattern. On the hexagonal lattice this pattern resembles a lizard skin. The properties of CA are specified by the most simple majority rule that adapts selected cell state to the most frequent state of cells in its surrounding. The method could be of interest for manufacturing of textiles as well as for modeling of patterns on skin of various animals.

*Keywords:* *random field transformation, cellular automata, labyrinthine pattern, lizard skin.*

## 1. INTRODUCTION

Many technological problems require physical characterization of developing fields by a quantitative field variable $S(r,t)$ denoting measureable properties like material composition, surface roughness, color, etc. This variable generally depends on the position $r$ and time $t$. The development of the field can be physically described by the evolution equation [1]:

$$\partial S(r,t) / \partial t = G(S(r_o, t); r_o \in O(r)) \qquad (1)$$

Here $O(r)$ indicates a properly selected surrounding of the point $r$ and $G(...)$ denotes a nonlinear field generator function that generally includes differential and integral operators. Eq. 1 has already been utilized in the study of various rather complex phenomena leading to formation of patterns in technical and natural environments [1-7]. Among them the generation of patterns by various reaction-diffusion processes in chemical reactors, plasma and biological environments are the most outstanding [2,7,8]. Fig.1 shows the field of ionization waves in a turbulent plasma developed from random initial conditions as determined by the numerical solution of a nonlinear integro-differential reaction-diffusion equation of type Eq. 1. Surprisingly, it resembles a characteristic pattern of a muscular structure.

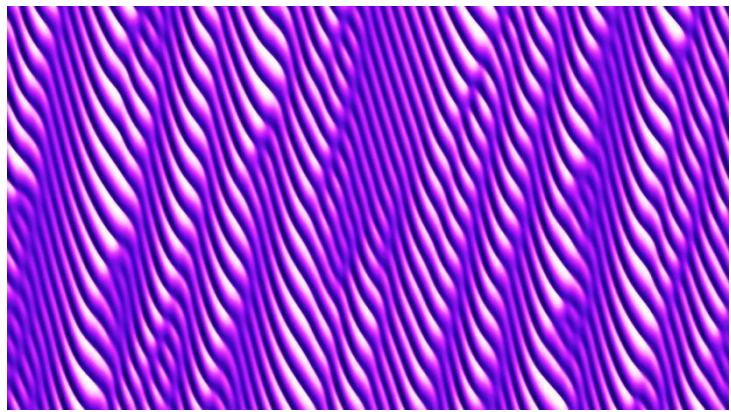

**Figure 1:** Distribution of turbulent ionization waves in a plasma of a glow discharge in a tube of argon. The distribution was determined by solving a nonlinear integro-differential reaction-diffusion equation of type Eq. 1 using random initial conditions and periodic boundary conditions [2]. The horizontal coordinate corresponds to time and the vertical to the axial coordinate of the discharge tube.

For the numerical treatment a discrete joint variable $s = (r,t)$ is usually introduced into evolution equation (1) and then it is transformed into the mapping relation [2,4]:

$$S(s) = G(S(s_o); s_o \in O(s)) \qquad (2)$$

This relation maps the old field distribution into the new one: $S(s_o) \rightarrow S(s)$. In such a case Eq. 2 represents a general form of multi-dimensional cellular automaton (CA) [8]. In order to apply it, one has to specify the generator function $G(\ldots)$, a discrete lattice of cells, initial and boundary conditions, as well as to solve Eq. 2 numerically. The generator function $G(\ldots)$ can be specified either theoretically, based upon physical laws governing the treated phenomenon [8], or empirically from experimental records of the field $S(s)$ [4,6]. In the last case a neural network can be applied to express the generator function in terms of recorded data [1,3,4,5]. Eq. 2 is convenient for modeling of patterns in graphic art and technology as well as in biology [1-7].

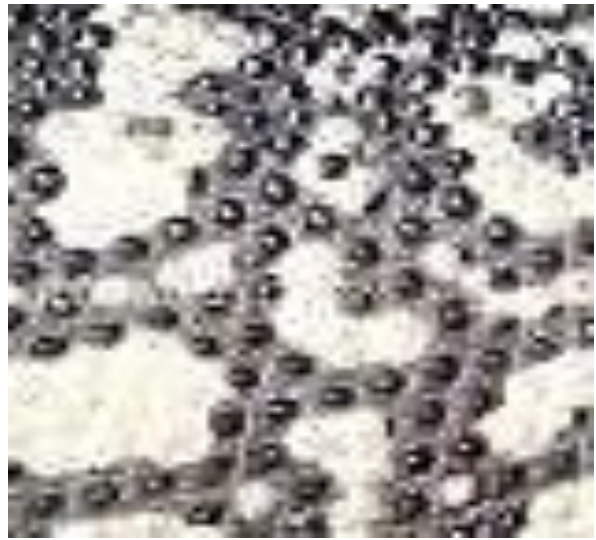

**Figure 2:** A characteristic pattern on a lizard skin.

## 2. SPECIFICATION OF CELLULAR AUTOMATON

The goal of the present article is to demonstrate applicability of Eq. 2 by modeling patterns of complex irregular form. For this purpose we first consider transformation of an initially random field by a two-dimensional cellular automaton into a new pattern resembling a lizard skin. This example is selected since it could be of interest for manufacturing of textiles as well as for modeling of patterns on skin of various animals [6]. For the specification of the CA structure we examine the pattern on a sample of lizard skin shown in Fig. 2. The pattern is comprised of scales representing approximately hexagonal cells, and consequently, we first accept a hexagonal lattice as the basis for the operation of our two-dimensional CA. In this case each cell is surrounded by six neighbor cells as shown in Fig.3.

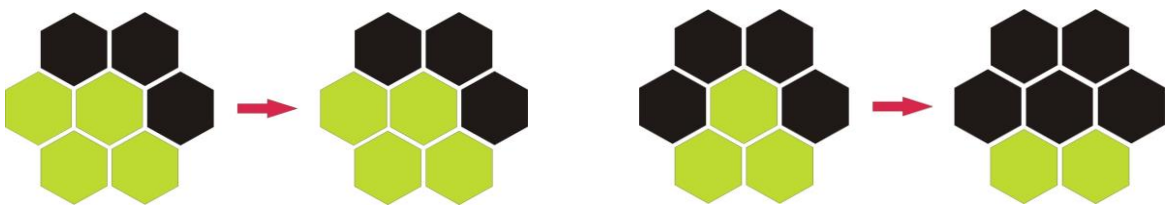

**Figure 3:** Two characteristic examples of center cell adaptation to its surrounding by the majority rule, left: without change, right: with a change of state.

## 3. EXAMPLES

In the first example we apply the hexagonal lattice. The initial random field distribution is shown in Fig. 4 **A** while the distribution after $1^{st}$, $2^{nd}$, and $10^{th}$ action are shown in Figs. 4 **B**, **C**, **D** respectively. In the first CA action many cells change the color, but the number of changes ΔN in a single action is decreasing with the number A of CA actions and ceases after several steps. This property is demonstrated in Fig. 5. It shows the total number of state changes N in dependence of the number A of CA actions. It is evident that our rule removes single jumps in the field distribution and yields more smooth variation in the final pattern. In spite of this smoothing effect the irregular character of the labyrinthine pattern is still preserved. This property does not depend on particular properties of the initial random field sample and resembles the effect of a digital filtering or changing of patterns by convolutional neural networks. Comparison of the final distribution in Fig. 4 **D** with the distribution on a real lizard skin shown in Fig. 2 reveals surprising similarity of characteristic features.

Similar properties as on the hexagonal lattice exhibits CA also on a quadratic lattice. Figs. 6 and 7 show results corresponding to Figs. 4 and 5. In this case the lines of the final pattern appear more cornered and the number of changes caused by CA actions is higher as in the hexagonal case. At some places a cyclic changing of state caused by successive CA actions can take place that leads to permanent increasing of N with A.

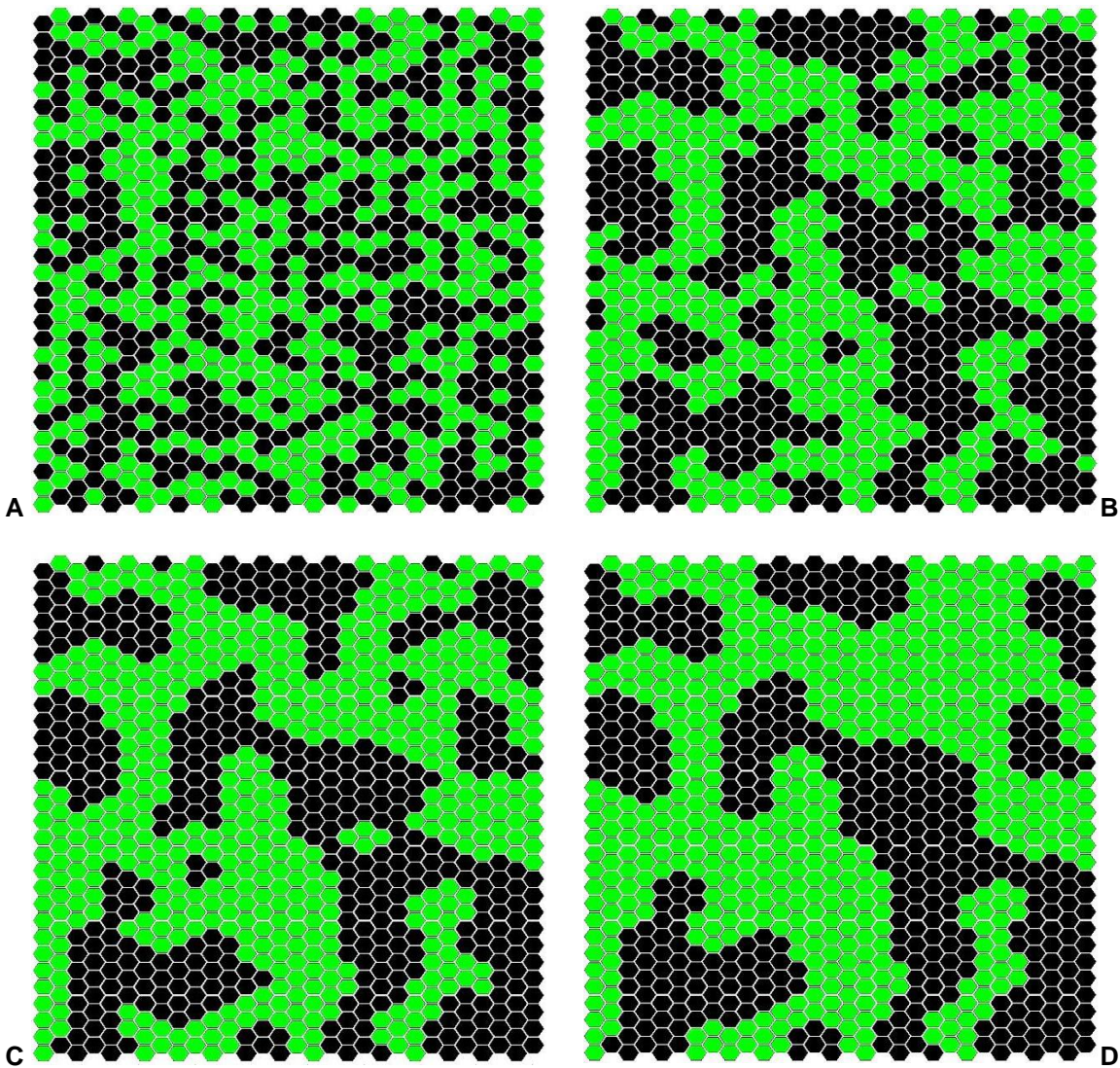

**Figure 4:** Distribution of the initial random field – **A** and distributions after $1^{st}$ - **B**, $2^{nd}$ - **C**, and $10^{th}$ – **D** action of CA on the hexagonal lattice.

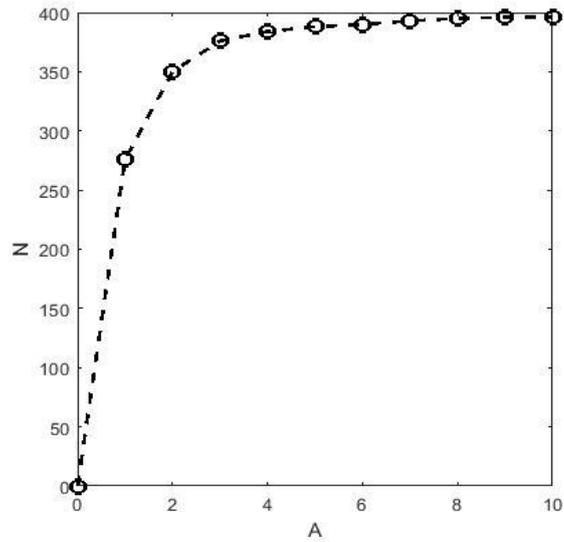

**Figure 5:** Number of changes N caused by A actions of CA on the hexagonal lattice.

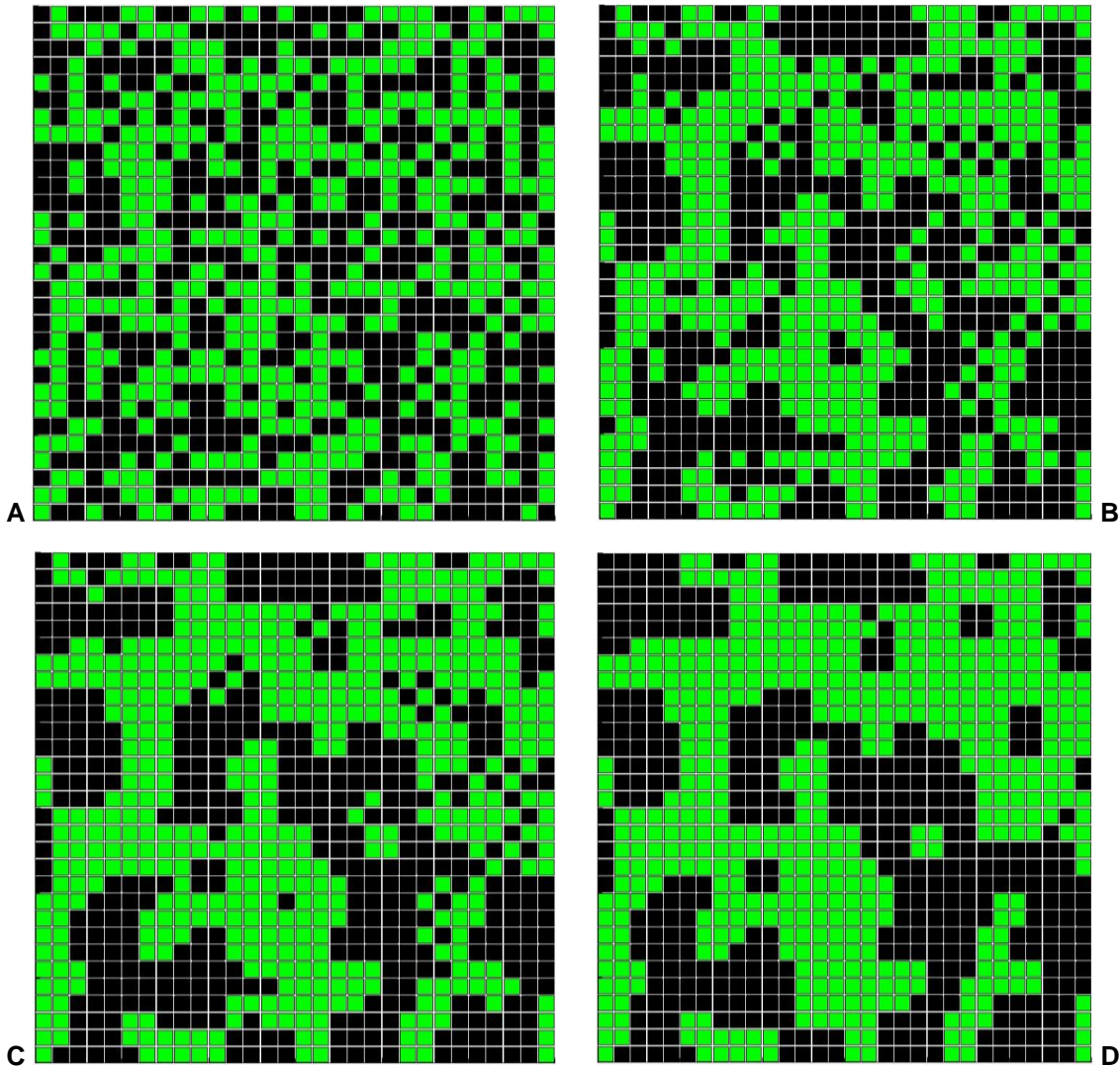

**Figure 6:** Distribution of the initial random field – **A** and distributions after 1st - **B**, 2nd - **C**, and 10th – **D** action of CA on the quadratic lattice.

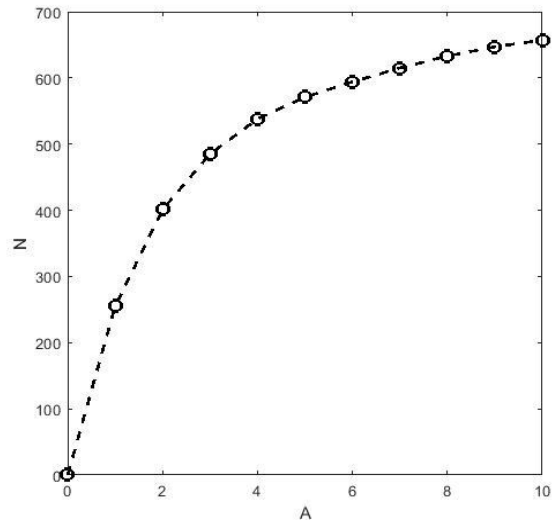

**Figure 7:** Number of changes N caused by A actions of CA on the quadratic lattice.

### 4. CONCLUSIONS

Our first example indicates applicability of the hexagonal cellular automaton for modeling of irregular complex patterns resembling the lizard skin. The second example reveals that the structure of CA lattice influences the roughness of the lines in the generated pattern. Since we consider just the nearest cells on the hexagonal or quadratic lattice, the lines of the generated patterns are rather cornered. By taking into account also more distant cells this weakness can be improved. At the specification of the generating function **G**(**…**) we here apply just the most simple majority rule. However, by changing this function, as well as the basic lattice, the properties of patterns can be well adapted to given samples [1,4,5]. For this purpose the function **G**(**…**) can be determined from given field records as has been described elsewhere [4,6].

**Corresponding author:**
Lovrenc ŠVEGL
Faculty of Natural Sciences and Engineering
Home: Poklukarjeva 66
SI-1000, Ljubljana, Slovenia
phone: +070691383
e-mail: lovrenc.svegl@gmail.com

**Co-author:**
Igor GRABEC
Faculty of Mechanical Engineering;
Aškerčeva cesta 6
1000 Ljubljana, Slovenia
phone: 041678182
email: igor.grabec@fs.uni-lj.si